\documentclass[aps,twocolumn,showpacs,prb,floatfix,superscriptaddress]{revtex4-1}

\usepackage{graphicx}
\usepackage{color}
\usepackage{tabularx}
\usepackage{epsfig}
\usepackage{amsmath}
\usepackage{amssymb}
\usepackage{graphicx}
\usepackage{dcolumn}
\usepackage{bm}
\usepackage{wasysym}
\usepackage{times}

\definecolor{grn}{rgb}{0,0,0.54}

\newcommand{\bra}[1]{\ensuremath{\langle #1 |}}
\newcommand{\ket}[1]{\ensuremath{| #1 \rangle}}
\newcommand{\expect}[1]{\langle \rangle}

\usepackage[colorlinks=true, urlcolor=blue, linkcolor=red, citecolor=blue, pdftex]{hyperref}
\hypersetup{pdfstartview={FitV}, pdfpagemode = UseNone, pdfpagelayout=SinglePage, bookmarksnumbered={true}}

\begin{document}


\title{Critical Correlations for Short-Range Valence-Bond Wave Functions on the Square Lattice}

\author{A. Fabricio Albuquerque}
\author{Fabien Alet}
\affiliation{Laboratoire de Physique Th{\' e}orique, Universit{\' e} de Toulouse and CNRS, UPS (IRSAMC),
F-31062 Toulouse, France}


\date{\today}
\pacs{75.10.Kt,75.10.Jm,75.40.Mg}

\begin{abstract}
We investigate the arguably simplest $SU(2)$-invariant wave functions capable of accounting for spin-liquid behavior,
expressed in terms of nearest-neighbor valence-bond states on the square lattice and characterized
by different topological invariants. While such wave-functions are known to exhibit short-range spin
correlations, we perform Monte Carlo simulations and show that four-point correlations decay algebraically
with an exponent $1.16(4)$. This is reminiscent of the {\it classical} dimer problem, albeit with a slower decay. Furthermore,
these correlators are found to be spatially modulated according to a wave-vector related to the topological
invariants. We conclude that a recently proposed spin Hamiltonian that stabilizes the here considered wave-function(s) as its
(degenerate) ground-state(s) should exhibit gapped spin and gapless non-magnetic excitations.\end{abstract}

\maketitle


{\it Introduction ---} The quest for quantum spin-liquid (QSL) states of matter\cite{balents:10} is a longstanding research topic
that can be traced back to Anderson's proposal.\cite{anderson:73} Building on earlier work, he conjectured
that strong quantum fluctuations, enhanced by frustration and/or low coordination, would weaken $SU(2)$-broken order and occasionally
drive an antiferromagnet towards a ``disordered" state with exponentially decaying spin correlations, describable in terms of short-ranged
spin-singlet, or {\em valence-bond} (VB), degrees of freedom.\cite{anderson:73,fazekas:74} Interest in Anderson's insight was further
triggered in connection with the cuprates since, soon after their discovery, spin-singlets in a QSL were interpreted as ``pre-formed Cooper
pairs" that would superconduct upon doping.\cite{anderson:87}

Major advances (reviewed in Refs.~\onlinecite{balents:10,misguich:10,misguich:05}) have taken place since the original
proposal,\cite{anderson:73} including a classification of possible QSL states\cite{wen:02} and explicit realizations in lattice
models in dimension $d>1$.\cite{misguich:99,balents:02,hermele:04a,Fujimoto:05,raman:05,seidel:09,kitaev:06a,yao:07,meng:10}
Also, theoretical ideas put forward in the early days of high-$T_{\rm c}$\cite{liang:88,sutherland:88} have been considerably
developed and resulted in a full-fledged formalism\cite{beach:06} as well as efficient numerical approaches\cite{lou:07} for
handling VB states. On the experimental side, a number of compounds have been shown not to display magnetic order down to
the lowest accessible temperatures,\cite{balents:10} much below the energy scale set by exchange
interactions, and are thus candidates for the realization of QSLs. However, in spite of such advances, a complete characterization
of QSL states is still missing, precluding unambiguous identification of experimental realizations, since absence of magnetic order
does not exclude the occurrence of, for instance, more conventional valence-bond crystals (VBC) that break lattice symmetries (see
{\it e.g.}~Ref.~\onlinecite{mambrini:06}).

Within this context, we investigate a family of nearest-neighbor VB (NN-VB) states on the square lattice by performing Monte
Carlo (MC) simulations based on a recently introduced algorithm.\cite{sandvik:10} We revisit the pioneering work by
Sutherland,\cite{sutherland:88} where a closely related NN-VB state was introduced, and provide a thorough characterization
of the arguably simplest $SU(2)$-invariant wave functions capable of accounting for QSL behavior. Although it has long been known
that NN-VB states on the square lattice are non-magnetic,\cite{liang:88} the possibility of other types of order, such as VBC, has not
yet been excluded. Despite their simplicity, and consequent theoretical appeal, the NN-VB states possess highly
non-trivial properties, that we explore in what follows.

{\it Wave functions ---} NN-VB states are obtained by contracting spins attached to NN sites $i$ and $j$ of a lattice into a
singlet state, $[i,j]=\frac{1}{\sqrt{2}}(\ket{\! \! \uparrow_{i} \downarrow_{j}}-\ket{\! \! \downarrow_{i} \uparrow_{j}})$.
Since each spin only pairs with one of its neighbors at a time, there is a one-to-one correspondence between NN-VB configurations
and those of hard-core {\em classical} dimers on the same lattice.\cite{kasteleyn:61,temperley:61,fisher:61,fisher:63}

\begin{figure}
\includegraphics*[width=0.225\textwidth,angle=270]{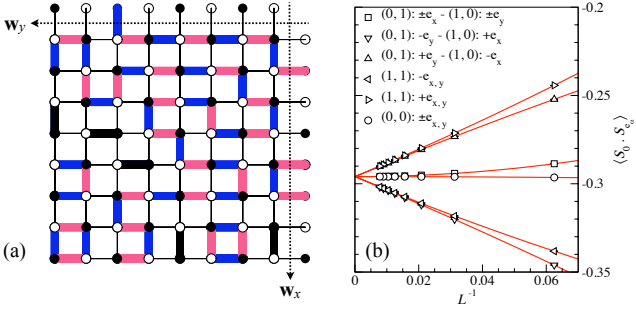}
\caption{(Color online) (a) Transition graph between two NN-VB configurations on the square lattice (sublattices
                ${\mathcal A}$/${\mathcal B}$ are indicated by filled/open circles). Reference lines for the winding numbers
                $\mathbf{w}=(w_{x},w_{y})$ are indicated by dashed lines: configurations with $\mathbf{w} =(1,0)$ (red lines)
                and $\mathbf{w}=(0,1)$ (blue lines) are shown. Trivial loops with coinciding VBs are depicted as thick-black
                lines and a loop winding in both directions is evident. (b) NN spin correlations versus $L^{-1}$ (from MC
                simulations) for winding sectors $\mathbf{w}=(0,0)$, $(0,1)$, $(1,0)$ and $(1,1)$. For ${\mathbf{w}} \neq (0,0)$,
                correlations along $\pm {\mathbf e}_{x}$ and $\pm {\mathbf e}_{y}$ are discriminated. Lines are linear-quadratic fits.}
\label{fig:Wnd}
\end{figure}

A crucial property of VB states is their non-orthogonality. The overlap between two VB configurations is given
by ${\langle \psi_1 | \psi_2 \rangle} = \pm 2^{N_{\mathcal{L}} - \frac{N}{2}}$,\cite{sutherland:88} where $N=L^{2}$
is the number of sites and $N_{\mathcal L}$ the number of loops in the transition graph obtained by superposing the
dimer configurations associated to $\ket{\psi_1}$ and $\ket{\psi_2}$ [Fig.~\ref{fig:Wnd}(a)]. For the square and other bipartite
lattices, that can be split into two sublattices ${\mathcal A}$ and ${\mathcal B}$, overlaps between arbitrary VB states can be
ensured to be always {\em positive}, so that stochastic methods apply (see below), by choosing $i \in {\mathcal A}$
and $j \in {\mathcal B}$ in the anti-symmetric singlet $[i,j]$.

An additional important point concerns the fact that dimer configurations can be split into {\em topological
sectors}.\cite{misguich:05} On a torus, the transition graph for two dimer coverings belonging to different topological
sectors displays non-local loops that wind around the system [Fig.~\ref{fig:Wnd}(a)], so that one dimer configuration can not be
continuously deformed onto the other via {\em local} dimer rearrangements. For bipartite lattices the number of topological sectors is
extensive and each sector can be labelled by topological invariants termed {\em winding numbers}, ${\mathbf{w}}=(w_x, w_y)$: $w_x$
($w_y$) is defined as the difference between the number of ${\mathcal B} \leftarrow {\mathcal A}$ and ${\mathcal A} \rightarrow
{\mathcal B}$ dimers along a reference line in the $y$ ($x$) direction [Fig.~\ref{fig:Wnd}(a)]. VB configurations characterized by
{\em different} ${\mathbf{w}}$ are {\em orthogonal} in the thermodynamic limit: the transition graph between two such configurations,
$\ket{c_{\mathbf{w_1}}}$ and $\ket{c_{\mathbf{w_2}}}$, contains at least one winding loop that comprises a minimum of $L$ dimers,
so that $N_{\mathcal L} \leq (N-L)/2$, implying that ${\langle c_{\mathbf{w_1}} | c_{\mathbf{w_2}} \rangle} \leq 2^{- \frac{L}{2}}$ and
vanishes when $L \rightarrow \infty$.\cite{bonesteel:89}

Having introduced the ingredients, we are able to write down the NN-VB wave functions we wish to investigate:
\begin{equation}
\ket{\psi_{\mathbf{w}}}=\sum_{c_{\mathbf{w}}} \ket{c_{\mathbf{w}}}~.
\label{eq:Psi}
\end{equation}
In contrast to the wave function analyzed by Sutherland,\cite{sutherland:88} who did not take the existence of topological
sectors into account and considered an equal amplitude superposition of {\em all} NN-VB states, each $\ket{\psi_{\mathbf{w}}}$
is an equal amplitude superposition of VB configurations $\ket{c_{\mathbf{w}}}$ with {\em fixed winding numbers} ${\mathbf{w}}=
(w_x, w_y)$. Our motivation for doing so is our previous remark that ${\langle \psi_{\mathbf{w_1}} | \psi_{\mathbf{w_2}} \rangle} = 0$
for $\mathbf{w_1} \neq \mathbf{w_2}$ in the thermodynamic limit. Interestingly, this implies that each $\ket{\psi_{\mathbf{w}}}$
is a (degenerate) ground-state on a torus of the {\em local} spin Hamiltonian recently proposed by Cano and Fendley.\cite{cano:10}
Although the number of winding sectors is extensive on a torus, we will show that it is possible to infer the properties of arbitrary states
$\ket{\psi_{\mathbf{w}}}$ by focusing on a few sectors with low $\mathbf{w}$.

{\it Algorithms ---} The expectation value of an observable ${\mathcal O}$ in Eq.~(\ref{eq:Psi}) can be measured by evaluating
\begin{equation}
\langle {\mathcal O} \rangle_{\mathbf{w}} =
\frac{1}{Z} \sum_{c_{\mathbf{w},1},c_{\mathbf{w},2}} \frac{\bra{c_{\mathbf{w},1}} {\cal O} \ket{c_{\mathbf{w},2}}}{\langle c_{\mathbf{w},1}
| c_{\mathbf{w},2} \rangle} \langle c_{\mathbf{w},1} | c_{\mathbf{w},2} \rangle~,
\label{eq:aver}
\end{equation}
where $Z= {\langle \psi_{\mathbf{w}} | \psi_{\mathbf{w}} \rangle}$ is the normalization. As first pointed out in Ref.~\onlinecite{liang:88},
$\langle {\mathcal O} \rangle_{\mathbf{w}}$ can be efficiently computed in a stochastic manner: the estimator
${\bra{c_{\mathbf{w},1}} {\cal O} \ket{c_{\mathbf{w},2}}}/{\langle c_{\mathbf{w},1} | c_{\mathbf{w},2} \rangle}$, that for most
observables of interest is readily evaluated by analyzing the loop structure in the transition graphs,\cite{beach:06} is sampled by generating
pairs of NN-VB configurations $\ket{c_{\mathbf{w},1}}$ and $\ket{c_{\mathbf{w},2}}$ according to the statistical weight given by their overlap
${\langle c_{\mathbf{w},1} | c_{\mathbf{w},2} \rangle} =2^{N_{\mathcal{L}}({\mathbf{w}};1,2) - \frac{N}{2}}$
[$N_{\mathcal{L}}({\mathbf{w}};1,2)$ denotes the number of loops in the transition graph ${\langle c_{\mathbf{w},1} | c_{\mathbf{w},2} \rangle}$].

Major advances in sampling techniques have been achieved since the work by Liang {\em et al.}~\cite{liang:88} and particularly well
suited to our purposes is a recently introduced algorithm.\cite{sandvik:10} Basically (we refer to Ref.~\onlinecite{sandvik:10}
for details), one combines non-local updates for the underlying dimer configurations, so to ensure small auto-correlations times, with spin updates 
that allow for an efficient sampling of the overlap weight, with unitary acceptance rate. By relying on this algorithm, we simulate systems with
periodic boundary conditions (PBC) of sizes of up to $L=128$. Topological symmetry is easily implemented in the simulations
by starting from a configuration in a given winding sector and discarding measurements for all MC moves that change ${\mathbf{w}}$: for
large $L$ winding updates are exponentially rare and this only causes small efficiency losses.

\begin{figure}
\includegraphics*[width=0.45\textwidth]{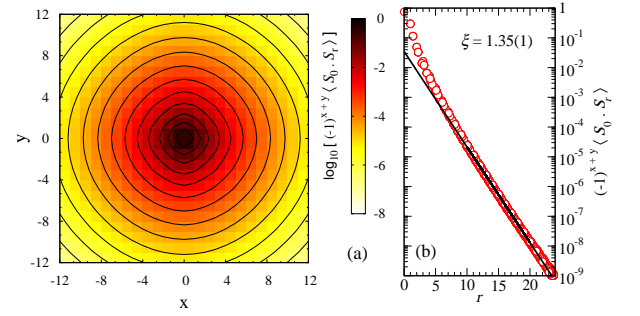}
\caption{(Color online) (a) Spin correlation $(-1)^{\mathbf r} \langle {\mathbf S}_{{\mathbf 0}} \cdot {\mathbf S}_{{\mathbf r}} \rangle$ between the spin
                at the origin and spins located at ${\mathbf r} = x{\mathbf e}_x + y{\mathbf e}_y$. (b) $(-1)^{\mathbf r}\langle {\mathbf S}_{{\mathbf 0}} \cdot
                {\mathbf S}_{{\mathbf r}} \rangle$ versus distance $r$. An exponential fit, $(-1)^{\mathbf r} \langle {\mathbf S}_{{\mathbf 0}} \cdot
                {\mathbf S}_{{\mathbf r}} \rangle \sim \exp(-r/\xi)$, yields the correlation length $\xi=1.35(1)$. Data for $L=128$ and $\mathbf{w}=(0,0)$.}
\label{fig:spin}
\end{figure}

{\it Short-range spin order ---} We start by analyzing the spin texture in wavefunction Eq.~(\ref{eq:Psi}). NN spin correlations,
$\langle {\mathbf S}_{{\mathbf r}} \cdot {\mathbf S}_{{\mathbf r}\pm{\mathbf e}_\alpha} \rangle$ (${\mathbf e}_\alpha$ is the unit vector
in the $\alpha=x,y$ direction) are plotted as a function of inverse system size $L^{-1}$ in Fig.~\ref{fig:Wnd}(b) for
${\mathbf{w}}=(0,0)$, $(1,0)$, $(0,1)$ and $(1,1)$. Deviations among results for different ${\mathbf{w}}$ are observed for small
systems and, additionally, for ${\mathbf{w}} \neq (0,0)$ vertical/horizontal and ${\mathcal A} \rightarrow {\mathcal B}$ /
${\mathcal B} \rightarrow {\mathcal A}$ correlations differ. However, all data converge to the same value in the
thermodynamic limit, according to a linear-quadratic best-fit analysis. Although we have no rigorous
justification for such scaling, we obtain $\langle {\mathbf S}_{{\mathbf r}} \cdot {\mathbf S}_{{\mathbf r}+ {\mathbf e}_\alpha}
\rangle = -0.295953(7)$ when $L \rightarrow \infty$ for all sectors. Spin correlations $\langle {\mathbf S}_{{\mathbf 0}} \cdot
{\mathbf S}_{{\mathbf r}} \rangle$ as a function of distance are plotted in Fig.~\ref{fig:spin}(a-b) for $L=128$ and $\mathbf{w}=(0,0)$
(virtually identical results are obtained for other $L$ and $\mathbf{w}$) and display a perfect staggered pattern consistent with
(isotropic) short-range N\'{e}el order. $\langle {\mathbf S}_{{\mathbf 0}} \cdot {\mathbf S}_{{\mathbf r}} \rangle$ decays very fast with
$|{\mathbf r}|$ and an exponential fit [Fig.~\ref{fig:spin}(b)] yields $\xi=1.35(1)$ for the spin correlation length.

\begin{figure}
\includegraphics*[width=0.425\textwidth,angle=270]{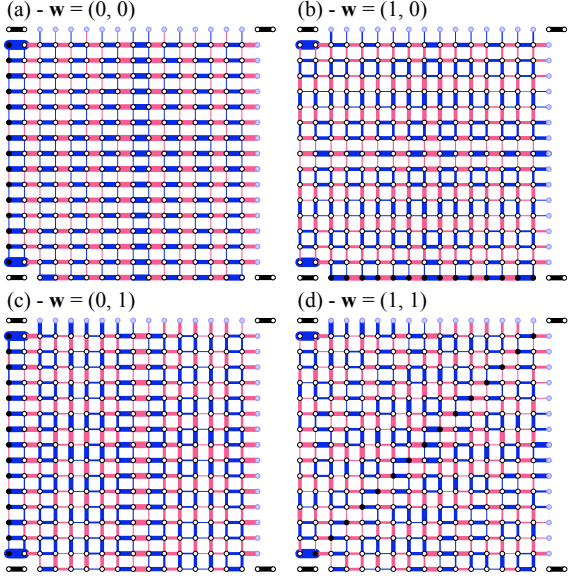}
\caption{(Color online) MC results for $rC^{ijkl}$, for $L=16$ (PBC) and indicated ${\mathbf{w}}$. In all panels, the reference bond
                is indicated by a thick-black line and the thickness of the remaining ones is proportional to $rC^{ijkl}$: blue (pale-red) lines
                indicate positive (negative) values. Lines along which $C_{\parallel} ^{ijkl}$ is strongest are indicated by black circles.}
\label{fig:spatial}
\end{figure}
\begin{figure}
\begin{center}
\includegraphics*[width=0.22\textwidth,angle=270]{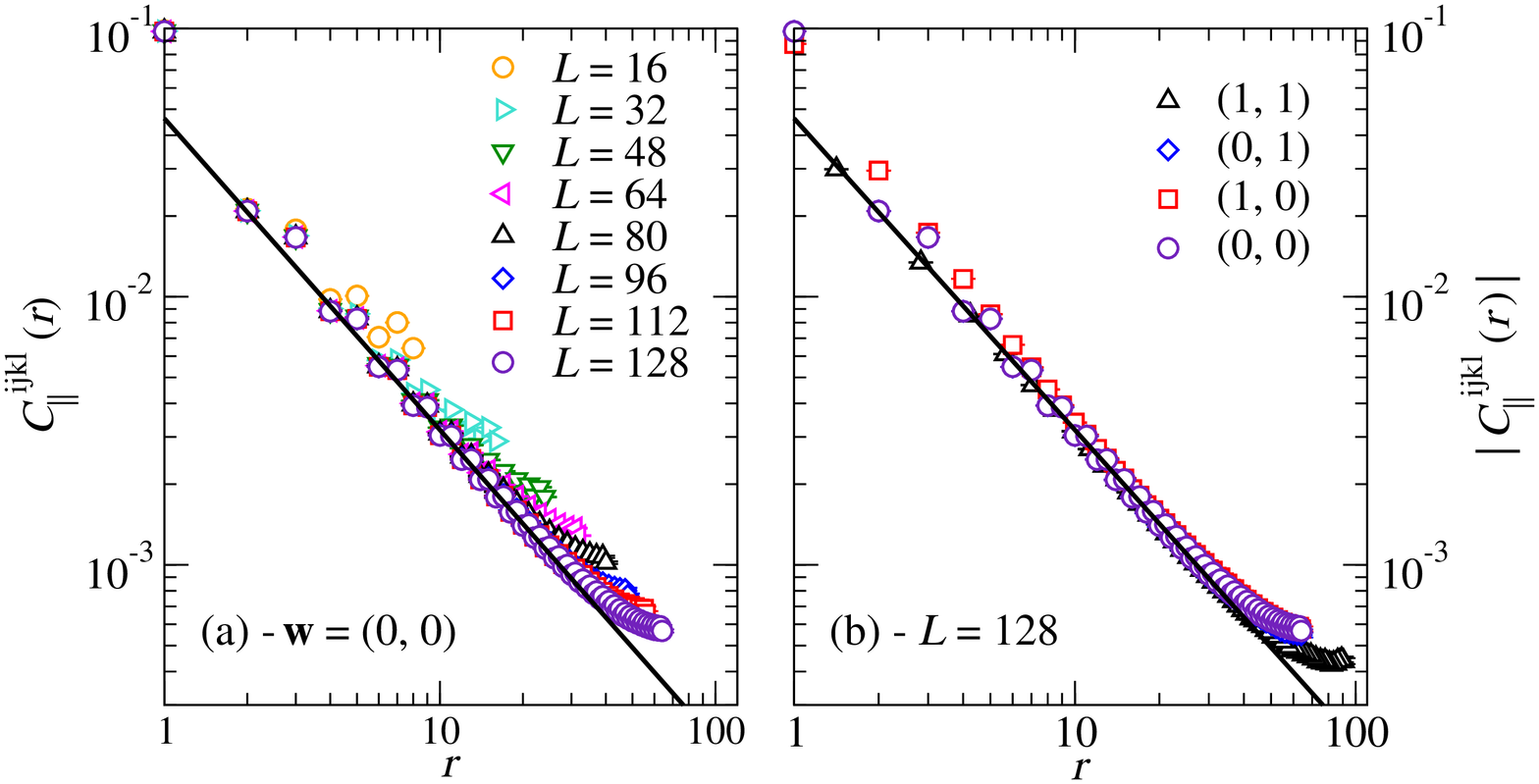}
 \vspace{0.3cm}
\includegraphics*[width=0.22\textwidth,angle=270]{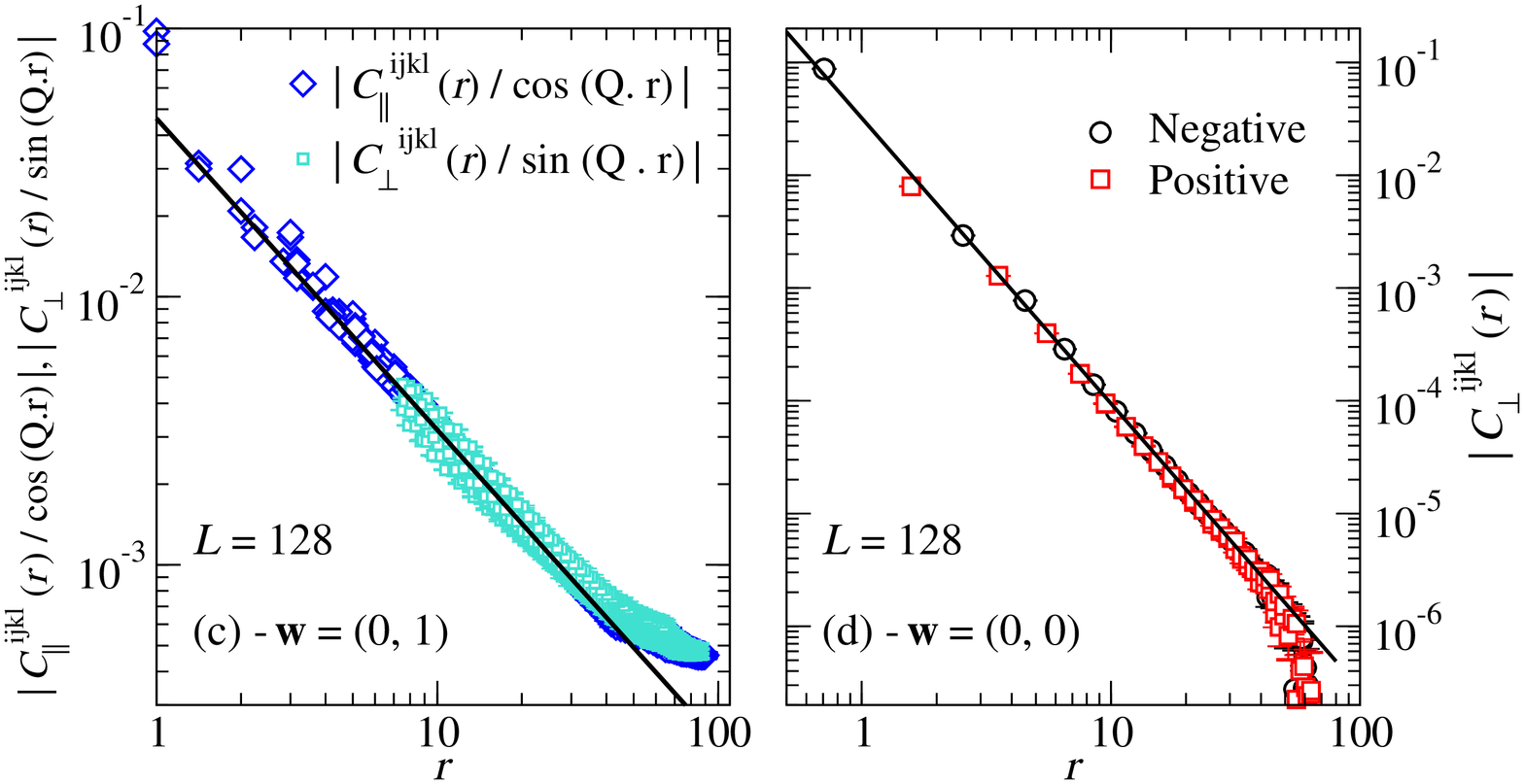}
\caption{(Color online) MC results for $C^{ijkl}$ as a function of distance. (a) Longitudinal correlations
                $C_{\parallel}^{ijkl}$ for ${\mathbf{w}}=(0,0)$ and various system sizes. (b) $C_{\parallel}^{ijkl}$ along zero-phase
                anti-nodal directions (filled circles in Fig.~\ref{fig:spatial}) for $L=128$ and ${\mathbf{w}}=(0,0)$, $(1,0)$,
                $(0,1)$ and $(1,1)$. (c) $C_{\parallel}^{ijkl}(r) / \cos (\mathbf{Q} \cdot \mathbf{r})$ and $C_{\perp}^{ijkl} (r) / \sin (\mathbf{Q} \cdot \mathbf{r})$
                for ${\mathbf{w}}=(0,1)$ and $L=128$ (data separated from a nodal line by a parallel displacement $dx < 8$ are excluded).
               (d) Absolute value of transverse correlations $C_{\perp}^{ijkl}$ for ${\mathbf{w}}=(0,0)$ and $L=128$, along the line highlighted in
               Fig.~\ref{fig:spatial}(a). In (a-c) the line indicates our best fit yielding the exponent $\alpha=1.16(4)$ and in (d)
               the fit yielding $\alpha'=2.53(5)$.
                }
\label{fig:dimcorr}
\end{center}
\end{figure}

{\it Critical correlations ---} We proceed to the characterization of ``dimer order" by analyzing the four-point connected correlators
$C^{ijkl}=\langle ({\mathbf S}_{i}\cdot {\mathbf S}_{j}) ({\mathbf S}_{k}\cdot {\mathbf S}_{l}) \rangle -\langle {\mathbf S}_{i}\cdot
{\mathbf S}_{j} \rangle \langle {\mathbf S}_{k}\cdot {\mathbf S}_{l}\rangle$, where both $i,j$ and $k,l$ are NN sites on the square
lattice. In Fig.~\ref{fig:spatial} we show MC data for the spatial dependence of $rC^{ijkl}$ ($r$ is the distance between dimers) for
$L=16$ and sectors ${\mathbf{w}}=(0,0)$, $(1,0)$, $(0,1)$ and $(1,1)$. We first notice that both $C_{\parallel}^{ijkl}$
(correlations for parallel dimers $i$, $j$ and $k$, $l$) and $C_{\perp}^{ijkl}$ (perpendicular dimers $i$, $j$ and $k$, $l$) are spatially
modulated for ${\mathbf{w}} \neq (0,0)$. Inspection of the results in Fig.~\ref{fig:spatial}, and similar ones for higher ${\mathbf{w}}$
(not shown), allows us to deduce that modulation for $C_{\parallel}^{ijkl}$ [$C_{\perp}^{ijkl}$] is entirely accounted for by a phase
factor $\cos (\mathbf{Q} \cdot \mathbf{r})$ [$\sin (\mathbf{Q} \cdot \mathbf{r})$], with a wave-vector given in terms of the winding
numbers, $\mathbf{Q}= \frac{2\pi}{L}(w_y, w_x)$. This inference is confirmed by our quantitative analysis below.\cite{classical}
Furthermore, we notice that no clear spatial dependence is noticeable for $rC_{\parallel}^{ijkl}$ in Fig.~\ref{fig:spatial}(a), suggesting
that four-point correlations in Eq.~(\ref{eq:Psi}) decay algebraically with $r$ with an exponent close to unity (see below).

In Fig.~\ref{fig:dimcorr}(a-b) we plot $C_{\parallel}^{ijkl}(r)$ along the anti-nodal
lines with strongest correlations (hence smallest relative errors), highlighted in Fig.~\ref{fig:spatial}, for: (a) ${\mathbf{w}} =(0,0)$
and various system sizes $L$ and (b) $L=128$ and ${\mathbf{w}}=(0,0)$, $(1,0)$, $(0,1)$ and $(1,1)$. Results are consistent with
power-law decay, $C_{\parallel}^{ijkl} \sim r^{-\alpha}$, as conjectured in Refs.~\onlinecite{sutherland:88,tasaki:89}.
Fitting the data in Fig.~\ref{fig:dimcorr}(b) we arrive at $\alpha=1.16(4)$. Although small deviations from
algebraic behavior are seen for large distances in Fig.~\ref{fig:dimcorr}(a-c), the value of $r$ at which they start to occur
increases linearly with $L$ (not shown) and we thus conclude that this ``upturn" in Fig.~\ref{fig:dimcorr}(a-c) is merely a
finite-size effect.

We analyze the spatial modulations for correlations and in Fig.~\ref{fig:dimcorr}(c) we plot $C_{\parallel}^{ijkl}(r)$ and $C_{\perp} ^{ijkl}(r)$
for $L=128$ and ${\mathbf{w}}=(0,1)$. By respectively dividing $C_{\parallel}^{ijkl}(r)$, $C_{\perp} ^{ijkl}(r)$ by the phase factors
$\cos (Q_x dx)$, $\sin (Q_x dx)$ [$dx=0$ along the anti-nodal line for $C_{\parallel}^{ijkl}(r)$ highlighted in Fig.~\ref{fig:spatial}(c) and
$\mathbf{Q}= \frac{2\pi}{L}(1, 0)$ in this case], we notice that all curves collapse, confirming that $C_{\parallel}^{ijkl}(r)$ and $C_{\perp} ^{ijkl}(r)$
decay with the same exponent and are indeed modulated according to the phase factors $\cos (\mathbf{Q} \cdot \mathbf{r})$ and $\sin
(\mathbf{Q} \cdot \mathbf{r})$, with $\mathbf{Q}= \frac{2\pi}{L}(w_y, w_x)$.

Sub-leading corrections to the scaling exponent can be obtained by analyzing $C_{\perp}^{ijkl}$ in the ${\mathbf{w}}=(0,0)$ sector
[see Fig.~\ref{fig:spatial}(a)].  Data for $C_{\perp}^{ijkl}$ are plotted as a function of $r$ in Fig.~\ref{fig:dimcorr}(d), for $L=128$ and
${\mathbf{w}}=(0,0)$. A fit yields $\alpha'=2.53(5)$ for the sub-leading exponent.

\begin{figure}
\includegraphics*[width=0.225\textwidth,angle=90]{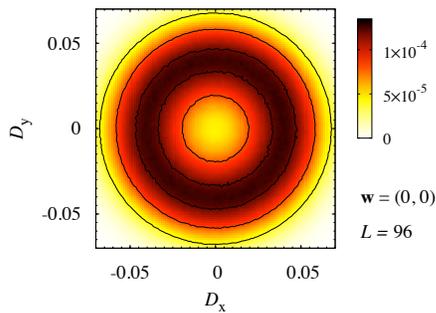}
\caption{(Color online) Histogram $P(D_x,D_y)$ for $L=96$ and ${\mathbf{w}}=(0,0)$.}
\label{fig:histogram}
\end{figure}

Finally, we address the point of what specific type of quasi-long-range dimer order is encoded in Eq.~(\ref{eq:Psi}). In doing so,
we analyze the dimer order parameter ${\mathbf D}$ defined by $D_{\alpha} = N^{-1} \sum_{\mathbf r} (-1)^{{\mathbf r}_\alpha}
{\mathbf S}_{\mathbf r}\cdot {\mathbf S}_{{\mathbf r}+{\mathbf e}_\alpha}$.
Due to the absence of long-range order, $\langle {\bf D} \rangle$ is expected to vanish when $L \rightarrow \infty$. However, information
concerning the symmetry of the quasi-ordered state is obtainable by analyzing the angular dependence on $\phi = \arctan({D_{y}/D_{x}})$
in the histogram $P(D_x,D_y)$ for occurrences of $D_x$ and $D_y$ in the simulations. In Fig.~\ref{fig:histogram} we plot $P(D_x,D_y)$
for $L=96$ and ${\mathbf{w}}=(0,0)$. Commonly observed VBCs on the square lattice, ``columnar" and ``plaquette" states (for a detailed
account see Ref.~\onlinecite{mambrini:06}), would display, respectively, peaks located at $\phi=\{0, \pm \pi/2,\pi\}$ and $\phi=\{\pm \pi/4,\pm 3 \pi/4\}$.
However, no angular structure is evident in Fig.~\ref{fig:histogram} and data are in favor of an continuous $U(1)$ symmetry, {\em a priori} different
from the $U(1)$ symmetry associated to topological degeneracy.

{\it Conclusions ---} We have investigated NN-VB wave functions on the square lattice [Eq.~(\ref{eq:Psi})],
characterized by winding numbers ${\mathbf{w}}$, by performing MC simulations. We confirm earlier findings
in favor of short-ranged spin order\cite{liang:88,beach:07} and, more interestingly, we find that dimer-dimer correlations are {\em critical},
a situation reminiscent of the one encountered for {\em classical} dimers~\cite{fisher:63} and thus for the ground-state of the quantum
dimer model (QDM) on the square lattice at the Rokhsar-Kivelson point.\cite{rokhsar:88} However, such correlations decay considerably
slower for Eq.~(\ref{eq:Psi}) than in the classical case, suggesting increased tendency towards VBC order: an exponent $\alpha= 1.16(4)$
accounts for the decay of both longitudinal and transverse correlation for {\em all} ${\mathbf{w}}$ studied, while in the classical case one has
$\alpha_{\rm class.}=2$.\cite{fisher:63} In this context, it would be interesting to analyze how exponents evolve by
considering the overlap as a tunable parameter,\cite{rokhsar:88} between the herein studied case and the limit of orthogonal dimer
configurations of the QDM.\cite{rokhsar:88}

From a broader perspective, we analyze how the wave-function Eq.~(\ref{eq:Psi}) fits into the general classification of QSL states.\cite{wen:02}
While we stress that our work concerns {\em wave-functions} and not the full spectrum of a given Hamiltonian, we notice that each
$\ket{\psi_{\mathbf{w}}}$ [Eq.~(\ref{eq:Psi})] is a (degenerate) ground-state of the local model of Ref.~\onlinecite{cano:10} on a torus.
We thus predict such $SU(2)$-invariant Hamiltonian to display {\it gapped spin} excitations, due to short-ranged spin order,\cite{sutherland:88,
kohmoto:88} and {\it gapless non-magnetic excitations}, since four-point correlations are critical and a theorem by Hastings applies.\cite{hastings:04}
Adopting the terminology of Ref.~\onlinecite{wen:02}, the latter excitations correspond to gapless gauge modes. Given its extensive degeneracy on a torus, our results altogether suggest that the ground-state of the model of Ref.~\onlinecite{cano:10} is a gapped $U(1)$ or $SU(2)$ spin liquid,\cite{wen:02} a state believed to be
{\em unstable} for a generic local spin model. In this context, the complete characterization of the spectrum of the model of Ref.~\onlinecite{cano:10}
and of perturbations thereof would be of high interest. 

Directions for further research opened up by our work include the study of wave-functions similar to Eq.~(\ref{eq:Psi}) in
different geometries. In $d=3$, we expect NN-VB states on bipartite lattices to display spin order, as it happens for the simple cubic
lattice,\cite{beach:07} but it would be interesting to search for traces of the Coulomb phase of dimer models in $d=3$.\cite{huse:03} NN-VB
wave-functions on $d=2$ geometrically frustrated lattices also deserve investigation.\cite{raman:05,seidel:09} However, a {\em sign
problem} precludes MC simulations as we perform here and alternative approaches are called for in this case.

{\it Note Added ---} While preparing this manuscript we became aware of related work by Tang {\em et al.}.\cite{tang:10}

{\it Acknowledgments ---} We thank P.~Fendley, M.~Mambrini and G.~Misguich for useful discussions. Our MC codes are based upon the
ALPS libraries.\cite{troyer:98,albuquerque:07} This work was performed using HPC resources from GENCI-CCRT
(Grant 2010-x2010050225) and CALMIP, and is supported by the French ANR program ANR-08-JCJC-0056-01.

\bibliographystyle{apsrev4-1}

\begin{thebibliography}{38}%
\makeatletter
\providecommand \@ifxundefined [1]{%
 \@ifx{#1\undefined}
}%
\providecommand \@ifnum [1]{%
 \ifnum #1\expandafter \@firstoftwo
 \else \expandafter \@secondoftwo
 \fi
}%
\providecommand \@ifx [1]{%
 \ifx #1\expandafter \@firstoftwo
 \else \expandafter \@secondoftwo
 \fi
}%
\providecommand \natexlab [1]{#1}%
\providecommand \enquote  [1]{``#1''}%
\providecommand \bibnamefont  [1]{#1}%
\providecommand \bibfnamefont [1]{#1}%
\providecommand \citenamefont [1]{#1}%
\providecommand \href@noop [0]{\@secondoftwo}%
\providecommand \href [0]{\begingroup \@sanitize@url \@href}%
\providecommand \@href[1]{\@@startlink{#1}\@@href}%
\providecommand \@@href[1]{\endgroup#1\@@endlink}%
\providecommand \@sanitize@url [0]{\catcode `\\12\catcode `\$12\catcode
  `\&12\catcode `\#12\catcode `\^12\catcode `\_12\catcode `\%12\relax}%
\providecommand \@@startlink[1]{}%
\providecommand \@@endlink[0]{}%
\providecommand \url  [0]{\begingroup\@sanitize@url \@url }%
\providecommand \@url [1]{\endgroup\@href {#1}{\urlprefix }}%
\providecommand \urlprefix  [0]{URL }%
\providecommand \Eprint [0]{\href }%
\@ifxundefined \urlstyle {%
  \providecommand \doi  [0]{\begingroup \@sanitize@url \@doi}%
  \providecommand \@doi [1]{\endgroup \@@startlink {\doibase
  #1}doi:\discretionary {}{}{}#1\@@endlink }%
}{%
  \providecommand \doi  [0]{doi:\discretionary{}{}{}\begingroup
  \urlstyle{rm}\Url }%
}%
\providecommand \doibase [0]{http://dx.doi.org/}%
\providecommand \Doi [0]{\begingroup \@sanitize@url \@Doi }%
\providecommand \@Doi  [1]{\endgroup\@@startlink{\doibase#1}\@@Doi}%
\providecommand \@@Doi [1]{#1\@@endlink}%
\providecommand \selectlanguage [0]{\@gobble}%
\providecommand \bibinfo  [0]{\@secondoftwo}%
\providecommand \bibfield  [0]{\@secondoftwo}%
\providecommand \translation [1]{[#1]}%
\providecommand \BibitemOpen [0]{}%
\providecommand \bibitemStop [0]{}%
\providecommand \bibitemNoStop [0]{.\EOS\space}%
\providecommand \EOS [0]{\spacefactor3000\relax}%
\providecommand \BibitemShut  [1]{\csname bibitem#1\endcsname}%
\bibitem [{\citenamefont {Balents}(2010)}]{balents:10}%
  \BibitemOpen
  \bibfield  {author} {\bibinfo {author} {\bibfnamefont {L.}~\bibnamefont
  {Balents}},\ }\href@noop {} {\bibfield  {journal} {\bibinfo  {journal}
  {Nature},\ }\textbf {\bibinfo {volume} {464}},\ \bibinfo {pages} {199}
  (\bibinfo {year} {2010})}\BibitemShut {NoStop}%
\bibitem [{\citenamefont {{Anderson}}(1973)}]{anderson:73}%
  \BibitemOpen
  \bibfield  {author} {\bibinfo {author} {\bibfnamefont {P.~W.}\ \bibnamefont
  {{Anderson}}},\ }\href@noop {} {\bibfield  {journal} {\bibinfo  {journal}
  {Mater. Res. Bull.},\ }\textbf {\bibinfo {volume} {8}},\ \bibinfo {pages}
  {153} (\bibinfo {year} {1973})}\BibitemShut {NoStop}%
\bibitem [{\citenamefont {{Fazekas}}\ and\ \citenamefont
  {{Anderson}}(1974)}]{fazekas:74}%
  \BibitemOpen
  \bibfield  {author} {\bibinfo {author} {\bibfnamefont {P.}~\bibnamefont
  {{Fazekas}}}\ and\ \bibinfo {author} {\bibfnamefont {P.~W.}\ \bibnamefont
  {{Anderson}}},\ }\href@noop {} {\bibfield  {journal} {\bibinfo  {journal}
  {Philos.~Mag.},\ }\textbf {\bibinfo {volume} {30}},\ \bibinfo {pages} {474}
  (\bibinfo {year} {1974})}\BibitemShut {NoStop}%
\bibitem [{\citenamefont {{Anderson}}(1987)}]{anderson:87}%
  \BibitemOpen
  \bibfield  {author} {\bibinfo {author} {\bibfnamefont {P.~W.}\ \bibnamefont
  {{Anderson}}},\ }\href@noop {} {\bibfield  {journal} {\bibinfo  {journal}
  {Science},\ }\textbf {\bibinfo {volume} {235}},\ \bibinfo {pages} {1196}
  (\bibinfo {year} {1987})}\BibitemShut {NoStop}%
\bibitem [{\citenamefont {Misguich}(2010)}]{misguich:10}%
  \BibitemOpen
  \bibfield  {author} {\bibinfo {author} {\bibfnamefont {G.}~\bibnamefont
  {Misguich}},\ }in\ \href@noop {} {\emph {\bibinfo {booktitle} {Exact Methods
  in Low-dimensional Statistical Physics and Quantum Computing}}},\ \bibinfo
  {editor} {edited by\ \bibinfo {editor} {\bibfnamefont {J.}~\bibnamefont
  {{Jacobsen}}} \bibnamefont{{\em et~al.}}}\ (\bibinfo
  {publisher} {Oxford Univ.~Press},\ \bibinfo {address} {Oxford},\ \bibinfo
  {year} {2010})\BibitemShut {NoStop}%
\bibitem [{\citenamefont {Misguich}\ and\ \citenamefont
  {Lhuillier}(2005)}]{misguich:05}%
  \BibitemOpen
  \bibfield  {author} {\bibinfo {author} {\bibfnamefont {G.}~\bibnamefont
  {Misguich}}\ and\ \bibinfo {author} {\bibfnamefont {C.}~\bibnamefont
  {Lhuillier}},\ }in\ \href@noop {} {\emph {\bibinfo {booktitle} {Frustrated
  spin systems}}},\ \bibinfo {editor} {edited by\ \bibinfo {editor}
  {\bibfnamefont {H.~T.}\ \bibnamefont {{Diep}}}}\ (\bibinfo  {publisher}
  {World-Scientific},\ \bibinfo {address} {Singapore},\ \bibinfo {year}
  {2005})\BibitemShut {NoStop}%
\bibitem [{\citenamefont {Wen}(2002)}]{wen:02}%
  \BibitemOpen
  \bibfield  {author} {\bibinfo {author} {\bibfnamefont {X.-G.}\ \bibnamefont
  {Wen}},\ }\Doi {10.1103/PhysRevB.65.165113} {\bibfield  {journal} {\bibinfo
  {journal} {Phys. Rev. B},\ }\textbf {\bibinfo {volume} {65}},\ \bibinfo
  {pages} {165113} (\bibinfo {year} {2002})}\BibitemShut {NoStop}%
\bibitem [{\citenamefont {Misguich}\ \emph {et~al.}(1999)\citenamefont
  {Misguich}, \citenamefont {Lhuillier}, \citenamefont {Bernu},\ and\
  \citenamefont {Waldtmann}}]{misguich:99}%
  \BibitemOpen
  \bibfield  {author} {\bibinfo {author} {\bibfnamefont {G.}~\bibnamefont
  {Misguich}} \bibnamefont{{\em et~al.}} 
  ,\ }\Doi
  {10.1103/PhysRevB.60.1064} {\bibfield  {journal} {\bibinfo  {journal} {Phys.
  Rev. B},\ }\textbf {\bibinfo {volume} {60}},\ \bibinfo {pages} {1064}
  (\bibinfo {year} {1999})}\BibitemShut {NoStop}%
\bibitem [{\citenamefont {Balents}\ \emph {et~al.}(2002)\citenamefont
  {Balents}, \citenamefont {Fisher},\ and\ \citenamefont
  {Girvin}}]{balents:02}%
  \BibitemOpen
  \bibfield  {author} {\bibinfo {author} {\bibfnamefont {L.}~\bibnamefont
  {Balents}}, \bibinfo {author} {\bibfnamefont {M.~P.~A.}\ \bibnamefont
  {Fisher}}, \ and\ \bibinfo {author} {\bibfnamefont {S.~M.}\ \bibnamefont
  {Girvin}},\ }\Doi {10.1103/PhysRevB.65.224412} {\bibfield  {journal}
  {\bibinfo  {journal} {Phys. Rev. B},\ }\textbf {\bibinfo {volume} {65}},\
  \bibinfo {pages} {224412} (\bibinfo {year} {2002})}\BibitemShut {NoStop}%
\bibitem [{\citenamefont {Hermele}\ \emph {et~al.}(2004)\citenamefont
  {Hermele}, \citenamefont {Fisher},\ and\ \citenamefont
  {Balents}}]{hermele:04a}%
  \BibitemOpen
  \bibfield  {author} {\bibinfo {author} {\bibfnamefont {M.}~\bibnamefont
  {Hermele}}, \bibinfo {author} {\bibfnamefont {M.~P.~A.}\ \bibnamefont
  {Fisher}}, \ and\ \bibinfo {author} {\bibfnamefont {L.}~\bibnamefont
  {Balents}},\ }\Doi {10.1103/PhysRevB.69.064404} {\bibfield  {journal}
  {\bibinfo  {journal} {Phys. Rev. B},\ }\textbf {\bibinfo {volume} {69}},\
  \bibinfo {pages} {064404} (\bibinfo {year} {2004})}\BibitemShut {NoStop}%
\bibitem [{\citenamefont {Fujimoto}(2005)}]{Fujimoto:05}%
  \BibitemOpen
  \bibfield  {author} {\bibinfo {author} {\bibfnamefont {S.}~\bibnamefont
  {Fujimoto}},\ }\Doi {10.1103/PhysRevB.72.024429} {\bibfield  {journal}
  {\bibinfo  {journal} {Phys. Rev. B},\ }\textbf {\bibinfo {volume} {72}},\
  \bibinfo {pages} {024429} (\bibinfo {year} {2005})}\BibitemShut {NoStop}%
\bibitem [{\citenamefont {Raman}\ \emph {et~al.}(2005)\citenamefont {Raman},
  \citenamefont {Moessner},\ and\ \citenamefont {Sondhi}}]{raman:05}%
  \BibitemOpen
  \bibfield  {author} {\bibinfo {author} {\bibfnamefont {K.~S.}\ \bibnamefont
  {Raman}}, \bibinfo {author} {\bibfnamefont {R.}~\bibnamefont {Moessner}}, \
  and\ \bibinfo {author} {\bibfnamefont {S.~L.}\ \bibnamefont {Sondhi}},\ }\Doi
  {10.1103/PhysRevB.72.064413} {\bibfield  {journal} {\bibinfo  {journal}
  {Phys. Rev. B},\ }\textbf {\bibinfo {volume} {72}},\ \bibinfo {pages}
  {064413} (\bibinfo {year} {2005})}\BibitemShut {NoStop}%
\bibitem [{\citenamefont {Seidel}(2009)}]{seidel:09}%
  \BibitemOpen
  \bibfield  {author} {\bibinfo {author} {\bibfnamefont {A.}~\bibnamefont
  {Seidel}},\ }\Doi {10.1103/PhysRevB.80.165131} {\bibfield  {journal}
  {\bibinfo  {journal} {Phys. Rev. B},\ }\textbf {\bibinfo {volume} {80}},\
  \bibinfo {pages} {165131} (\bibinfo {year} {2009})}\BibitemShut {NoStop}%
\bibitem [{\citenamefont {{Kitaev}}(2006)}]{kitaev:06a}%
  \BibitemOpen
  \bibfield  {author} {\bibinfo {author} {\bibfnamefont {A.}~\bibnamefont
  {{Kitaev}}},\ }\href@noop {} {\bibfield  {journal} {\bibinfo  {journal} {Ann.
  Phys.},\ }\textbf {\bibinfo {volume} {321}},\ \bibinfo {pages} {2} (\bibinfo
  {year} {2006})}\BibitemShut {NoStop}%
\bibitem [{\citenamefont {Yao}\ and\ \citenamefont {Kivelson}(2007)}]{yao:07}%
  \BibitemOpen
  \bibfield  {author} {\bibinfo {author} {\bibfnamefont {H.}~\bibnamefont
  {Yao}}\ and\ \bibinfo {author} {\bibfnamefont {S.~A.}\ \bibnamefont
  {Kivelson}},\ }\Doi {10.1103/PhysRevLett.99.247203} {\bibfield  {journal}
  {\bibinfo  {journal} {Phys. Rev. Lett.},\ }\textbf {\bibinfo {volume} {99}},\
  \bibinfo {pages} {247203} (\bibinfo {year} {2007})}\BibitemShut {NoStop}%
\bibitem [{\citenamefont {{Meng}}\ \emph {et~al.}(2010)\citenamefont {{Meng}},
  \citenamefont {{Lang}}, \citenamefont {{Wessel}}, \citenamefont {{Assaad}},\
  and\ \citenamefont {{Muramatsu}}}]{meng:10}%
  \BibitemOpen
  \bibfield  {author} {\bibinfo {author} {\bibfnamefont {Z.~Y.}\ \bibnamefont
  {{Meng}}} \bibnamefont{{\em et~al.}},\ }\href@noop {}
  {\bibfield  {journal} {\bibinfo  {journal} {Nature},\ }\textbf {\bibinfo
  {volume} {464}},\ \bibinfo {pages} {847} (\bibinfo {year}
  {2010})}\BibitemShut {NoStop}%
\bibitem [{\citenamefont {Liang}\ \emph {et~al.}(1988)\citenamefont {Liang},
  \citenamefont {Dou\c{c}ot},\ and\ \citenamefont {Anderson}}]{liang:88}%
  \BibitemOpen
  \bibfield  {author} {\bibinfo {author} {\bibfnamefont {S.}~\bibnamefont
  {Liang}}, \bibinfo {author} {\bibfnamefont {B.}~\bibnamefont {Dou\c{c}ot}}, \
  and\ \bibinfo {author} {\bibfnamefont {P.~W.}\ \bibnamefont {Anderson}},\
  }\Doi {10.1103/PhysRevLett.61.365} {\bibfield  {journal} {\bibinfo  {journal}
  {Phys. Rev. Lett.},\ }\textbf {\bibinfo {volume} {61}},\ \bibinfo {pages}
  {365} (\bibinfo {year} {1988})}\BibitemShut {NoStop}%
\bibitem [{\citenamefont {Sutherland}(1988)}]{sutherland:88}%
  \BibitemOpen
  \bibfield  {author} {\bibinfo {author} {\bibfnamefont {B.}~\bibnamefont
  {Sutherland}},\ }\Doi {10.1103/PhysRevB.37.3786} {\bibfield  {journal}
  {\bibinfo  {journal} {Phys. Rev. B},\ }\textbf {\bibinfo {volume} {37}},\
  \bibinfo {pages} {3786} (\bibinfo {year} {1988})}\BibitemShut {NoStop}%
\bibitem [{\citenamefont {{Beach}}\ and\ \citenamefont
  {{Sandvik}}(2006)}]{beach:06}%
  \BibitemOpen
  \bibfield  {author} {\bibinfo {author} {\bibfnamefont {K.~S.~D.}\
  \bibnamefont {{Beach}}}\ and\ \bibinfo {author} {\bibfnamefont {A.~W.}\
  \bibnamefont {{Sandvik}}},\ }\href@noop {} {\bibfield  {journal} {\bibinfo
  {journal} {Nucl. Phys. B},\ }\textbf {\bibinfo {volume} {750}},\ \bibinfo
  {pages} {142} (\bibinfo {year} {2006})}\BibitemShut {NoStop}%
\bibitem [{\citenamefont {Lou}\ and\ \citenamefont {Sandvik}(2007)}]{lou:07}%
  \BibitemOpen
  \bibfield  {author} {\bibinfo {author} {\bibfnamefont {J.}~\bibnamefont
  {Lou}}\ and\ \bibinfo {author} {\bibfnamefont {A.~W.}\ \bibnamefont
  {Sandvik}},\ }\Doi {10.1103/PhysRevB.76.104432} {\bibfield  {journal}
  {\bibinfo  {journal} {Phys. Rev. B},\ }\textbf {\bibinfo {volume} {76}},\
  \bibinfo {pages} {104432} (\bibinfo {year} {2007})}\BibitemShut {NoStop}%
\bibitem [{\citenamefont {Mambrini}\ \emph {et~al.}(2006)\citenamefont
  {Mambrini}, \citenamefont {L\"auchli}, \citenamefont {Poilblanc},\ and\
  \citenamefont {Mila}}]{mambrini:06}%
  \BibitemOpen
  \bibfield  {author} {\bibinfo {author} {\bibfnamefont {M.}~\bibnamefont
  {Mambrini}} \bibnamefont{{\em et~al.}}
  ,\ }\Doi
  {10.1103/PhysRevB.74.144422} {\bibfield  {journal} {\bibinfo  {journal}
  {Phys. Rev. B},\ }\textbf {\bibinfo {volume} {74}},\ \bibinfo {pages}
  {144422} (\bibinfo {year} {2006})}\BibitemShut {NoStop}%
\bibitem [{\citenamefont {Sandvik}\ and\ \citenamefont
  {Evertz}(2010)}]{sandvik:10}%
  \BibitemOpen
  \bibfield  {author} {\bibinfo {author} {\bibfnamefont {A.~W.}\ \bibnamefont
  {Sandvik}}\ and\ \bibinfo {author} {\bibfnamefont {H.~G.}\ \bibnamefont
  {Evertz}},\ }\Doi {10.1103/PhysRevB.82.024407} {\bibfield  {journal}
  {\bibinfo  {journal} {Phys. Rev. B},\ }\textbf {\bibinfo {volume} {82}},\
  \bibinfo {pages} {024407} (\bibinfo {year} {2010})}\BibitemShut {NoStop}%
\bibitem [{\citenamefont {{Kasteleyn}}(1961)}]{kasteleyn:61}%
  \BibitemOpen
  \bibfield  {author} {\bibinfo {author} {\bibfnamefont {P.~W.}\ \bibnamefont
  {{Kasteleyn}}},\ }\href@noop {} {\bibfield  {journal} {\bibinfo  {journal}
  {Physica},\ }\textbf {\bibinfo {volume} {27}},\ \bibinfo {pages} {1209}
  (\bibinfo {year} {1961})}\BibitemShut {NoStop}%
\bibitem [{\citenamefont {{Temperley}}\ and\ \citenamefont
  {{Fisher}}(1961)}]{temperley:61}%
  \BibitemOpen
  \bibfield  {author} {\bibinfo {author} {\bibfnamefont {H.~N.~V.}\
  \bibnamefont {{Temperley}}}\ and\ \bibinfo {author} {\bibfnamefont {M.~E.}\
  \bibnamefont {{Fisher}}},\ }\href@noop {} {\bibfield  {journal} {\bibinfo
  {journal} {Philos. Mag.},\ }\textbf {\bibinfo {volume} {6}},\ \bibinfo
  {pages} {1061} (\bibinfo {year} {1961})}\BibitemShut {NoStop}%
\bibitem [{\citenamefont {Fisher}(1961)}]{fisher:61}%
  \BibitemOpen
  \bibfield  {author} {\bibinfo {author} {\bibfnamefont {M.~E.}\ \bibnamefont
  {Fisher}},\ }\Doi {10.1103/PhysRev.124.1664} {\bibfield  {journal} {\bibinfo
  {journal} {Phys. Rev.},\ }\textbf {\bibinfo {volume} {124}},\ \bibinfo
  {pages} {1664} (\bibinfo {year} {1961})}\BibitemShut {NoStop}%
\bibitem [{\citenamefont {Fisher}\ and\ \citenamefont
  {Stephenson}(1963)}]{fisher:63}%
  \BibitemOpen
  \bibfield  {author} {\bibinfo {author} {\bibfnamefont {M.~E.}\ \bibnamefont
  {Fisher}}\ and\ \bibinfo {author} {\bibfnamefont {J.}~\bibnamefont
  {Stephenson}},\ }\Doi {10.1103/PhysRev.132.1411} {\bibfield  {journal}
  {\bibinfo  {journal} {Phys. Rev.},\ }\textbf {\bibinfo {volume} {132}},\
  \bibinfo {pages} {1411} (\bibinfo {year} {1963})}\BibitemShut {NoStop}%
\bibitem [{\citenamefont {Bonesteel}(1989)}]{bonesteel:89}%
  \BibitemOpen
  \bibfield  {author} {\bibinfo {author} {\bibfnamefont {N.~E.}\ \bibnamefont
  {Bonesteel}},\ }\Doi {10.1103/PhysRevB.40.8954} {\bibfield  {journal}
  {\bibinfo  {journal} {Phys. Rev. B},\ }\textbf {\bibinfo {volume} {40}},\
  \bibinfo {pages} {8954} (\bibinfo {year} {1989})}\BibitemShut {NoStop}%
\bibitem [{\citenamefont {Cano}\ and\ \citenamefont {Fendley}(2010)}]{cano:10}%
  \BibitemOpen
  \bibfield  {author} {\bibinfo {author} {\bibfnamefont {J.}~\bibnamefont
  {Cano}}\ and\ \bibinfo {author} {\bibfnamefont {P.}~\bibnamefont {Fendley}},\
  }\Doi {10.1103/PhysRevLett.105.067205} {\bibfield  {journal} {\bibinfo
  {journal} {Phys. Rev. Lett.},\ }\textbf {\bibinfo {volume} {105}},\ \bibinfo
  {pages} {067205} (\bibinfo {year} {2010})}\BibitemShut {NoStop}%
\bibitem [{\citenamefont {{Beach}}(2007)}]{beach:07}%
  \BibitemOpen
  \bibfield  {author} {\bibinfo {author} {\bibfnamefont {K.~S.~D.}\
  \bibnamefont {{Beach}}},\ }\href@noop {} { (\bibinfo {year} {2007})},\
  \bibinfo {note} {{cond-mat: 0707.0297}}\BibitemShut {NoStop}%
\bibitem [{cla()}]{classical}%
  \BibitemOpen
  \href@noop {} {}\bibinfo {note} {Similar effects are seen in correlators for
  classical dimers with fixed ${\mathbf{w}}$, yet with a different spatial
  structure (unpublished).}\BibitemShut {Stop}%
\bibitem [{\citenamefont {Tasaki}(1989)}]{tasaki:89}%
  \BibitemOpen
  \bibfield  {author} {\bibinfo {author} {\bibfnamefont {H.}~\bibnamefont
  {Tasaki}},\ }\Doi {10.1103/PhysRevB.40.9183} {\bibfield  {journal} {\bibinfo
  {journal} {Phys. Rev. B},\ }\textbf {\bibinfo {volume} {40}},\ \bibinfo
  {pages} {9183} (\bibinfo {year} {1989})}\BibitemShut {NoStop}%
\bibitem [{\citenamefont {{Rokhsar}}\ and\ \citenamefont
  {{Kivelson}}(1988)}]{rokhsar:88}%
  \BibitemOpen
  \bibfield  {author} {\bibinfo {author} {\bibfnamefont {D.~S.}\ \bibnamefont
  {{Rokhsar}}}\ and\ \bibinfo {author} {\bibfnamefont {S.~A.}\ \bibnamefont
  {{Kivelson}}},\ }\href@noop {} {\bibfield  {journal} {\bibinfo  {journal}
  {Phys. Rev. Lett.},\ }\textbf {\bibinfo {volume} {61}},\ \bibinfo {pages}
  {2376} (\bibinfo {year} {1988})}\BibitemShut {NoStop}%
\bibitem [{\citenamefont {Kohmoto}\ and\ \citenamefont
  {Shapir}(1988)}]{kohmoto:88}%
  \BibitemOpen
  \bibfield  {author} {\bibinfo {author} {\bibfnamefont {M.}~\bibnamefont
  {Kohmoto}}\ and\ \bibinfo {author} {\bibfnamefont {Y.}~\bibnamefont
  {Shapir}},\ }\Doi {10.1103/PhysRevB.37.9439} {\bibfield  {journal} {\bibinfo
  {journal} {Phys. Rev. B},\ }\textbf {\bibinfo {volume} {37}},\ \bibinfo
  {pages} {9439} (\bibinfo {year} {1988})}\BibitemShut {NoStop}%
\bibitem [{\citenamefont {Hastings}(2004)}]{hastings:04}%
  \BibitemOpen
  \bibfield  {author} {\bibinfo {author} {\bibfnamefont {M.~B.}\ \bibnamefont
  {Hastings}},\ }\Doi {10.1103/PhysRevB.69.104431} {\bibfield  {journal}
  {\bibinfo  {journal} {Phys. Rev. B},\ }\textbf {\bibinfo {volume} {69}},\
  \bibinfo {pages} {104431} (\bibinfo {year} {2004})}\BibitemShut {NoStop}%
\bibitem [{\citenamefont {Huse}\ \emph {et~al.}(2003)\citenamefont {Huse},
  \citenamefont {Krauth}, \citenamefont {Moessner},\ and\ \citenamefont
  {Sondhi}}]{huse:03}%
  \BibitemOpen
  \bibfield  {author} {\bibinfo {author} {\bibfnamefont {D.~A.}\ \bibnamefont
  {Huse}}, \bibinfo {author} {\bibfnamefont {W.}~\bibnamefont {Krauth}},
  \bibinfo {author} {\bibfnamefont {R.}~\bibnamefont {Moessner}}, \ and\
  \bibinfo {author} {\bibfnamefont {S.~L.}\ \bibnamefont {Sondhi}},\ }\Doi
  {10.1103/PhysRevLett.91.167004} {\bibfield  {journal} {\bibinfo  {journal}
  {Phys. Rev. Lett.},\ }\textbf {\bibinfo {volume} {91}},\ \bibinfo {pages}
  {167004} (\bibinfo {year} {2003})}\BibitemShut {NoStop}%
\bibitem [{\citenamefont {{Tang}}\ \emph {et~al.}(2010)\citenamefont {{Tang}},
  \citenamefont {{Sandvik}},\ and\ \citenamefont {{Henley}}}]{tang:10}%
  \BibitemOpen
  \bibfield  {author} {\bibinfo {author} {\bibfnamefont {Y.}~\bibnamefont
  {{Tang}}}, \bibinfo {author} {\bibfnamefont {A.~W.}\ \bibnamefont
  {{Sandvik}}}, \ and\ \bibinfo {author} {\bibfnamefont {C.~L.}\ \bibnamefont
  {{Henley}}},\ }\href@noop {} { (\bibinfo {year} {2010})},\ \bibinfo {note}
  {{ arXiv:1010:6146. Preliminary results were announced in: Y.~Tang and A.~W.~Sandvik, Bull.~Am.~Phys.~Soc.~{\bf 55},
  P38.12 (2010)}}\BibitemShut {NoStop}%
\bibitem [{\citenamefont {{Troyer}}\ \emph {et~al.}(1998)\citenamefont
  {{Troyer}}, \citenamefont {{Ammon}},\ and\ \citenamefont
  {{Heeb}}}]{troyer:98}%
  \BibitemOpen
  \bibfield  {author} {\bibinfo {author} {\bibfnamefont {M.}~\bibnamefont
  {{Troyer}}}, \bibinfo {author} {\bibfnamefont {B.}~\bibnamefont {{Ammon}}}, \
  and\ \bibinfo {author} {\bibfnamefont {E.}~\bibnamefont {{Heeb}}},\
  }\href@noop {} {\bibfield  {journal} {\bibinfo  {journal} {Lecture Notes in
  Comput. Sci.},\ }\textbf {\bibinfo {volume} {1505}},\ \bibinfo {pages} {191}
  (\bibinfo {year} {1998})}\BibitemShut {NoStop}%
\bibitem [{\citenamefont {{Albuquerque}}\ \emph {et~al.}(2007)\citenamefont
  {{Albuquerque}}, \citenamefont {{Alet}}, \citenamefont {{Dayal}},
  \citenamefont {{Feiguin}}, \citenamefont {{Fuchs}}, \citenamefont {{Gamper}},
  \citenamefont {{Gull}}, \citenamefont {{G\"urtler}}, \citenamefont
  {{Honecker}}, \citenamefont {{Igarashi}}, \citenamefont {{K\"orner}},
  \citenamefont {{Kozhevnikov}}, \citenamefont {{L\"auchli}}, \citenamefont
  {{Manmana}}, \citenamefont {{Matsumoto}}, \citenamefont {{McCulloch}},
  \citenamefont {{Michel}}, \citenamefont {{Noack}}, \citenamefont
  {{Pawlowski}}, \citenamefont {{Pollet}}, \citenamefont {{Pruschke}},
  \citenamefont {{Schollw\"ock}}, \citenamefont {{Todo}}, \citenamefont
  {{Trebst}}, \citenamefont {{Troyer}}, \citenamefont {{Werner}},\ and\
  \citenamefont {{Wessel}}}]{albuquerque:07}%
  \BibitemOpen
  \bibfield  {author} {\bibinfo {author} {\bibfnamefont {A.~F.}\ \bibnamefont
  {{Albuquerque}}} \bibnamefont{{\em et~al.}},\ }\href@noop {} {\bibfield  {journal}
  {\bibinfo  {journal} {J. Magn. Magn. Mater.},\ }\textbf {\bibinfo {volume}
  {310}},\ \bibinfo {pages} {1187} (\bibinfo {year} {2007})}\BibitemShut
  {NoStop}%
\end{thebibliography}

\end{document}